\documentclass[12pt]{article}
\usepackage{amsmath,cite}
\usepackage{amssymb, amsthm}
\newif\ifpdf
\ifx\pdfoutput\undefined
  \pdffalse       
\else
  \pdfoutput=1    
 \pdftrue
\fi

\ifpdf
  \usepackage[pdftex]{graphicx}
\else
  \usepackage[dvips,xdvi]{graphicx}
\fi
\usepackage{epsfig}
\usepackage{lscape}

\title{ Change of Measure in Midcurve Pricing}
\author{K.E. Feldman}
\date{}

\begin{document}

\maketitle
\begin{abstract} We derive measure change formulae required to price midcurve swaptions in the forward swap annuity measure with stochastic annuities' ratios. We construct the corresponding linear  and exponential terminal swap rate pricing models and show how they capture the midcurve swaption correlation skew.  
\end{abstract}

\newcommand{\docdate}{ \today \ }
\newcommand{\docsubject}{Unsmiling Midcurves Correlation  }
\newcommand{\dockeywords}{ }

\newcommand{\docabstract}{}
\renewcommand{\abstractname}{\vspace{-\baselineskip}}

\newcommand{\tl}{\tt}
\newcommand{\adj}{\text{adj}}
\newcommand{\mkt}{\text{mkt}}
\newcommand{\md}{\text{mod}}
\newcommand{\dd}{\text{d}}
\newcommand{\Prob}[2]{\text{Prob}^{#1}\left[#2\right]}

\newtheorem{Assumption}{Assumption}
\newtheorem{Prop}{Proposition}
\newtheorem{Th}{Theorem}
\newtheorem{Lemma}{Lemma}

\section*{Introduction}
An interest rate swap is a financial instrument with a triangle property. The value of two swaps $S_{t_1t_2}$, $S_{t_2t_3}$ between times $t_1$ and $t_2$ and between times $t_2$ and $t_3$ is equal to the value of the swap $S_{t_1t_3}$ between times $t_1$ and $t_3$ (we assume that all three swaps have the same fixed leg strike). Equivalently, we may say that the swap $S_{t_2t_3}$ is the difference between a long swap $S_{t_1t_3}$ and a short swap $S_{t_1t_2}$. To express views on swap rates in the future, the interest rate market actively trades options on swaps, i.e. swaptions. Swaptions are non-linear products. The triangle property of the swap generalises into the property of the swaptions by including the convexity. A portfolio of a vanilla swaption on the short swap $S_{t_1t_2}$ and an option (midcurve swaption) on the swap $S_{t_2t_3}$ is more expensive than the price of the long swaption $S_{t_1t_3}$ (when the strikes are the same, and the exercise time of all of the swaptions is the same, $t_1$ - the start of the short and long swaps).  If, in the Black-Scholes world, we assume also that the long and short annuities ratios to the annuity of the midcurve swap are deterministic, then from the swap triangle one can derive a useful  relationship for the volatilities of all of the three swap rates. The challenge comes when we look into the relations between the volatility smiles (skews) of those rates. In this paper, we discuss  a modelling approach  for pricing midcurve swaptions that allows one to take into account the stochasticity of the swaps' annuities and to generate pronounced correlation  skews which are typically observed in the midcurve swaption market.   

A midcurve swaption is an efficient  way to trade correlations between the short and long swap rates.  Others also used this product to trade on the difference between levels in the short and long term implied volatilities~\cite{StMcGraw}. Being the simplest product on forward volatility, midcurve swaptions can be used for the calibration of the mean reversion parameters in the one factor short rate models~\cite{AP2}.  

The rich structure of the interest rate market offers two approaches to modelling  the price of a midcurve swaption. The product can be viewed dynamically and be priced by modelling the time evolution of the underlying swap rate, or it can be viewed statically and its price can be derived from prices of closely related products - the long and the short swaptions traded in the market. 

We shall be looking at the static way of pricing the midcurve swaption using a generalisation of the triangle property of the swaps to the case of the swaptions. A midcurve swaption can be priced as an option on a weighted basket of the short and long swap rates with the same fixing date. The weights coefficients are functions of the swap annuities ratios. The industry standard is to freeze these ratios to be constants. Taking correlation as an input parameter, the weighted basket can be priced by pairing the short and long swap rate distributions via a copula.

Some use more advanced models to account for stochasticity of the annuity ratios. The approach that has been adopted by the larger banks is first to move both the short and the long swaps rates distributions to the same terminal (discount bond) measure, and then to approximate each of the short and long annuities by deterministic functions of the corresponding (short or long) swap rates. This is an extension of the idea~\cite{cedpit} where the authors developed a model that directly links constant maturity swap to volatilities of swaptions of all relevant tenors. Note that midcurve swaptions are not in the  scope of~\cite{cedpit}. This product is liquidly traded in the US Dollar market where the settlement style is physical. Thus, the natural pricing measure for this product is the annuity measure.  

While allowing a better risk management of the midcurve correlation skew, the terminal measure approach suffers from an inconsistency. In this paper, we show that once you fix the stochastic form of the annuity ratio, the measure change is no longer free. We derive the explicit formulae for the measure change in terms of the functional forms of the annuity ratios.  One other deficiency of the terminal measure approach are negative ratios of annuities. The exponential terminal swap rate model developed in this paper is free of this problem by construction. 

We analyse in detail the measure change formulae in the case where the annuity ratio is a linear or an exponential function of the short and the long swap rates. The price of a midcurve swaption is often parameterised by its implied correlation as a function of the strike. Even if we use a model that captures the implied volatility smiles of the long and short swap rates well, the implied correlation is still not a constant function of the strike. The termianl swap rate models with stochastic annuities developed in this paper give a handle to match the implied correlation skew. 

The effect studied in this paper is applicable in conjunction with any smile model. In particular, it is present in the flat volatility world.
We provide numerical results on how our methodology captures the midcurve correlation skew in the case when the underlying swap rates are modelled as standard normal variables with a flat smile (i.e. constant across all strikes' volatilities). 

\section{Product valuation}
A midcurve receiver swaption $W_{rec}=W_{rec}(S_{rec},T_{e(x)piry})$ on a swap $S_{rec}(T_{(s)tart},T_{(e)nd},K)$ with a fixed leg rate $K$  gives the holder an option to enter into a receiver swap $S_{rec}$ at expiry time $T_x$, where the swap starts on $T_s$, ends on $T_e$  and the holder receives the fixed rate $K$ accrued on a notional $N$ over all periods in the schedule formed by a sequence of dates:
$T^{fix}_1,\quad \dots,\quad T^{fix}_n=T_e$,  with $n$ payment dates in the fixed leg schedule. In return the holder pays floating rate payments on the sequence of dates from the floating rate schedule: $T^{fl}_1,\quad \dots,\quad T^{fl}_m=T_e$.
We will use short notations for the time intervals between two consecutive payments on each of the swap legs:
$\tau^{fix}_i=T^{fix}_i-T^{fix}_{i-1}$,  $i=1,\dots n$, $\tau^{fl}_j=T^{fl}_j-T^{fl}_{j-1}$,  $j=1,\dots m$,
$T^{fl}_0=T^{fix}_0=T_s$.

In order to price a swaption,  one uses the swap fixed leg annuity $A(t)$ ($t\le T_s$) as a numeraire:
\begin{equation}
A(t) = A(t,T_s,T_e)= \sum^n_{i=1}\tau^{fix}_i D(t,T^{fix}_i),
\label{eq3.1}
\end{equation}
where $D(t,T)$ is the relevant discount bond from $t$ to $T$.
We write the swap as:
\begin{eqnarray}
\ \ \ \ \ \ \  \ \ \ \ 
S_{rec}&=NA(t)(K-R(t))=NA(t,T_s,T_e)(K-R(t,T_s,T_e)),
\label{eq3.2}
\end{eqnarray}
where $R(t)=R(t,T_s,T_e)$ is the forward $T_s$-to-$T_e$-swap rate
 as seen at $t$.
Consequently, in order to price the  swaption $W_{rec}$, we can model the distribution for the $R(T_x,T_s,T_e)$ in the annuity measure and calculate the value of the swaption as:
\begin{eqnarray} 
W_{rec}(t)&=&A(t)\mathbb {E^A}[W(T_x)/A(T_x)]\nonumber\\
&=&A(t,T_s,T_e)N\mathbb {E^A}[[K-R(T_x,T_s,T_e)]^+],
\label{eq3.4}
\end{eqnarray}
where the superscript in $\mathbb {E^A}$ denotes the annuity measure with numeraire $A(t)$.

The distributions for $R(t_0,t_1,t_2)$ can be implied from the swaption market whenever $t_0 = t_1$.
 We are primarily interested in the distributions of the following two stochastic variables:
\begin{equation} 
R_s=R(T_x,T_x,T_s),\quad  R_e=R(T_x,T_x,T_e),
\label{eq4.1}
\end{equation}
where $R(T_x,T_x,T_s)$ and $R(T_x,T_x,T_e)$  are the swap rates of the corresponding "short" and "long" swaps.  
 Following~\cite{BL} and using~(\ref{eq3.4}) the probability density function, $\rm PDF$, for the distribution of the swap rate $R(T_x,T_x,T_J)$ with $J=s$ or $e$ in the corresponding annuity measure is given by
\begin{equation}
{\rm PDF}^J_{R_J}(r) = \frac{1}{A(t_0,T_x,T_J)\cdot N}\frac{\partial^2W_{rec}(t_0)}{\partial K^2}|_{K=r},
\label{eq5.2}
\end{equation}
where the derivative is taken with respect to the strike $K$ of the swaption 
\begin{equation}
W_{rec}(S_{rec}(T_x,T_J,K),T_x).
\label{eq5.3}
\end{equation}

Within the pricing approach provided by~(\ref{eq3.4}), the distributions for $R_s$, $R_e$ are specified in the corresponding annuity measures: $A(t,T_x,T_s)$, $A(t,T_x,T_e)$.
 The swap rate $R(T_x,T_s,T_e)$, that we are interested in, can be expressed as
\begin{align}
R(T_x,T_s,T_e)& =w_1\cdot R_e-w_2\cdot R_s,
\label{eq4.8}
\end{align}
where 
\begin{equation}
w_1=
\frac{A(T_x,T_x,T_e)}{A(T_x,T_s,T_e)}\quad
w_2=
\frac{A(T_x,T_x,T_s)}{A(T_x,T_s,T_e)}.
\label{eq4.9}
\end{equation}
Therefore, the stochastic variable $R(T_x,T_s,T_e)$ representing the underlying swap rate is a weighted difference of the stochastic variables representing the long and the short swap rates with stochastic coefficients. Note that the three stochastic variables 
\begin{equation} 
A_s=A(T_x,T_x,T_s),\quad  A_e=A(T_x,T_x,T_e),\quad and\quad
 A_u=A(T_x,T_s,T_e)
\label{eq4.10}
\end{equation}
are related via
\begin{equation} 
A(T_x,T_s,T_e) = A(T_x,T_x,T_e)-
  A(T_x,T_x,T_s).
\label{eq4.11}
\end{equation}
We are going to model the distribution of the swap rate $R(T_x,T_s,T_e)$ in terms of the distributions of $R(T_x,T_x,T_e)$, $R(T_x,T_x,T_s)$ and their correlation.  In order to do this we need to relate three annuity measures corresponding to $A(t,T_x,T_s)$, $A(t,T_x,T_e)$ and $A(t,T_s,T_e)$.

The Radon-Nikodym derivative for the measure change between  $A(t,T_x,T_J)$ measure, $J=s,e$, and $ A(t,T_s,T_e)$ measure can be reconstructed using the following identity:
\begin{eqnarray}
\ \ \ \ \ \ \ \
V(t_0)& =& A(t_0,T_x,T_J){\mathbb E^{A(T_x,T_J)}} \left[\frac{V(t)}{A(t,T_x,T_J)} \right]\nonumber\\
& =&  A(t_0,T_s,T_e){\mathbb E^{A(T_s,T_e)}} \left[\frac{V(t)}{A(t,T_s,T_e)} \right],
\label{eq5.6}
\end{eqnarray}
where $V(t)$ is the price of a traded security (which is a stochastic variable at any future time).
 Under the standard assumptions on attainable claims and measure changes,  equation~(\ref{eq5.6}) implies that for any stochastic process $X_t$, which is a function of the swap rate $R(T_x,T_x,T_J)$,
\begin{equation}
{\mathbb E^{A(T_s,T_e)}} \left[X_t \right] =  {\mathbb E^{A(T_x,T_J)}} \left[X_t \frac{A(t_0,T_x,T_J)}{A(t,T_x,T_J)} \cdot \frac{A(t,T_s,T_e)}{A(t_0,T_s,T_e)} \right].
\label{eq5.7}
\end{equation}
The quantity 
\begin{equation}
G_{J,T_x} =  \frac{A(T_x,T_s,T_e)}{A(T_x,T_x,T_J)}
\label{eq5.8}
\end{equation}
is itself a stochastic variable. We shall assume that it has a joint distribution with the swap rates $R(T_x,T_x,T_s)$  and  $R(T_x,T_x,T_e)$
\begin{equation}
{\rm PDF}^J_{G_J,R_s,R_e}(g_J,x,y),
\label{eq5.9}
\end{equation}
where the variable $g_J$ is used to indicate a stochastic value for $G_{J,T_x}$, the variable $x$ is used to indicate a stochastic value for $R(T_x,T_x,T_s)$, and  the variable $y$ is used to indicate a stochastic value for $R(T_x,T_x,T_e)$.

\begin{Lemma}
\label{mcf}
The measure change formulae for the marginals $\phi_s(x)$ and $\phi_e(y)$ of the joint distribution of the short and the long swap rates in the measure ($u$) associated  with the underlying swap annuity $A_u$ of the midcurve swaption are:
\begin{eqnarray}
\phi_s(x)&:=&{\rm PDF}^u_{R_s}(x) = \int^{+\infty}_{-\infty}\int^{+\infty}_{-\infty}{\rm PDF}^u_{R_u,R_s,R_e}(z,x,y)dzdy =\nonumber\\
&=&{\rm PDF}^s_{R_s}(x)\frac{A(t_0,T_x,T_s)}{A(t_0,T_s,T_e)} {\mathbb E^{A(T_x,T_s)}\left[G_{s,T_x}|R(T_x,T_x,T_s)=x\right]}
\label{eqmcf1}
\end{eqnarray} 
\begin{eqnarray}
\phi_e(y)&:=& {\rm PDF}^u_{R_e}(y)=\int^{+\infty}_{-\infty}\int^{+\infty}_{-\infty}{\rm PDF}^u_{R_u,R_s,R_e}(z,x,y)dzdx=\nonumber\\
&=&{\rm PDF}^e_{R_e}(y)\frac{A(t_0,T_x,T_e)}{A(t_0,T_s,T_e)} {\mathbb E^{A(T_x,T_e)}\left[G_{e,T_x}|R(T_x,T_x,T_e)=y\right]}
\label{eqmcf2}
\end{eqnarray}
\end{Lemma}
\noindent

\

Thus, given the PDFs of $R_e$ and $R_s$ in their natural (annuities) measures, we can derive the PDFs of $R_e$ and $R_s$ in the common $A(T_s,T_e)$-measure as soon as we can evaluate ${\mathbb E^{A(T_x,T_e)}\left[G_{e,T_x}|R(T_x,T_x,T_e)=y\right]}$, and
${\mathbb E^{A(T_x,T_s)}\left[G_{s,T_x}|R(T_x,T_x,T_s)=x\right]}$.

To evaluate the payoff of the midcurve swaption we will make an assumption that the stochastic variables $G_{J,T_x}$, $J=e,s$ from~(\ref{eq5.8}) are deterministic functions of the swap rates $R(T_x,T_x,T_s)$ and $R(T_x,T_x,T_e)$.  The integral formula for the payoff is:
\begin{align}
&\frac{W_{rec}(t_0)}{A(t_0,T_s,T_e)\cdot N} = {\mathbb E^{A(T_s,T_e)}}\left[[K-R(T_x,T_s,T_e)]^+\right]=\nonumber \\
&={\mathbb E^{A(T_s,T_e)}}\left[{\mathbb E^{A(T_s,T_e)}}\left[[K-R(t,T_s,T_e)]^+|R_s=x,R_e=y\right]\right]=\nonumber\\
&={\mathbb E^{A(T_s,T_e)}}\left[{\mathbb E^{A(T_s,T_e)}}\left[[K-\frac{A(T_x,T_x,T_e)}{A(T_x,T_s,T_e)}R_e+\frac{A(T_x,T_x,T_s)}{A(T_x,T_s,T_e)}R_s]^+|R_s=x,R_e=y\right]\right]\nonumber\\
&={\mathbb E^{A(T_s,T_e)}}\left[{\mathbb E^{A(T_s,T_e)}}\left[[K-w_1(y,x)y+w_2(y,x)x]^+|R_s=x,R_e=y\right]\right]=\nonumber\\
 \nonumber\\
&=\int^{+\infty}_{-\infty}\int^{+\infty}_{-\infty}{\mathbb E^{A(T_s,T_e)}}\left[[K-w_1(y,x)y+w_2(y,x)x]^+|R_s=x,R_e=y\right]\times\nonumber\\
&\times{\rm PDF}^u_{R_u,R_s,R_e}(z(x,y),x,y)dxdy=\nonumber\\
&=\int^{+\infty}_{-\infty}\int^{+\infty}_{-\infty}\left[K-w_1(y,x)y+w_2(y,x)x\right]^+
{\rm PDF}^u_{R_s,R_e}(x,y)dxdy,
\label{eq5.22}
\end{align}
where we omitted the symbol $z(x,y)$ in the last formula because it is fully determined by $x$ and $y$ due to our assumption on $G_{J,T_x}$, $J=e,s$.

In order to use~(\ref{eq5.22}) we need to specify the weight functions $w_1(y,x)$ and $w_2(y,x)$ as well as the full joint distribution ${\rm PDF}^u_{R_s,R_e}(x,y)$. The latter can be constructed using the copula technique applied to distributions $\phi_s(x)$ from~(\ref{eqmcf1}) and $\phi_e(y)$ from~(\ref{eqmcf2}).  
A popular choice is the Gaussian copula:
\begin{equation}
{\rm GC_{Join}}(x,y) = \phi(u,v,\rho) \frac{du}{dx} \frac{dv}{dy},
\label{eq5.30}
\end{equation}
where $\phi(u,v,\rho)$ is the joint normal PDF of two univariate normal variables with correlation $\rho$ and 
\begin{eqnarray}
\ \ \ \ \ \ \ \ \ \ \ \ \ \ \ \ \ 
u=\Phi^{-1}\left[cdf_s(x)\right],&\quad &v=\Phi^{-1}\left[cdf_e(y)\right],
\label{eq5.29}
\end{eqnarray}
where $\Phi$ is the CDF of a univariate normal variable and $cdf_s(x)$, $cdf_e(y)$ are the CDFs corresponding to pdfs $\phi_s(x)$ from~(\ref{eqmcf1}) and $\phi_e(y)$ from~(\ref{eqmcf2}).

\section{First order approximations}

The Radon-Nikodym derivative for measure change in~(\ref{eq5.7}) and the payoff in~(\ref{eq5.22})  depend only on the ratio of the annuities  $A(t,T_x,T_s)$ and $A(t,T_x,T_e)$. Therefore, to use the copula valuation by means of~(\ref{eq5.22})  it is sufficient to model dynamics of the ratio of annuities. A convenient way for modelling dynamics of the ratio of annuities is provided by the Terminal Swap Rate Model methodology. It covers the zero-th and the first order
 approximations for the ratio. We discuss the corresponding approximations below.

\

\noindent
{\bf Deterministic Annuity Ratio:} Assume that the conditional expectations in ~(\ref{eqmcf1}-\ref{eqmcf2}) are independent from the respective variables $x$ and $y$ (we may think of $G_{J,T_x}$, $J=e,s$ from~(\ref{eq5.8}), for example, as being deterministic). Then in~(\ref{eqmcf1}) $\phi_s(x)\equiv {\rm PDF}^s_{R_s}(x)$  and 
in~(\ref{eqmcf2}) $\phi_e(y)\equiv {\rm PDF}^e_{R_e}(y)$, i.e. no change of the measure is needed and
\begin{equation}
w_1(y,x)=G_{e,T_x}^{-1}=\frac{A(t_0,T_x,T_e)}{A(t_0,T_s,T_e)},\quad{}
w_2(y,x)=G_{s,T_x}^{-1}=\frac{A(t_0,T_x,T_s)}{A(t_0,T_s,T_e)}.
\label{eqN5.1}
\end{equation} 
This is exactly the constant annuity ratio assumption used in~\cite{AP3} for pricing midcurve swaptions by means of the Gaussian copula. 

\

\noindent
{\bf Linear Approximation:} We can approximate linearly the weights in~(\ref{eq5.22}) as
\begin{align}
&w_1(y,x)=G_{e,T_x}^{-1}=\nonumber\\
&=\frac{A(t_0,T_x,T_e)}{A(t_0,T_s,T_e)}(1+\mu_e(y-\hat R(t_0,T_x,T_e))+\mu_s(x-\hat R(t_0,T_x,T_s))),
\label{eqN5.2}
\end{align}
and
\begin{align}
&w_2(y,x)=G_{s,T_x}^{-1}=\nonumber\\
&=\frac{A(t_0,T_x,T_s)}{A(t_0,T_s,T_e)}(1+\nu_e(y-\hat R(t_0,T_x,T_e))+\nu_s(x-\hat R(t_0,T_x,T_s))),
\label{eqN5.3}
\end{align}
where $\hat R$ is used to underline that a measure change is needed to evaluate the corresponding quantity, so that 
\begin{eqnarray}
\ \ \ \ \ \ \ \ \ \ \ \ \ \ \ \ \ \
\hat R(t_0,T_x,T_e)&=&{\mathbb E^{A(T_s,T_e)}}\left[R(t,T_x,T_e)\right],\nonumber\\
\hat R(t_0,T_x,T_s)&=&{\mathbb E^{A(T_s,T_e)}}\left[R(t,T_x,T_s)\right],
\label{eqN5.4}
\end{eqnarray}
and both $w_1(y,x)$, $w_2(y,x)$ are $A(T_s,T_e)$-martingales.

Equating coefficients under $x$ and $y$ in $w_1(y,x)-w_2(y,x)=1$, we see that the four coefficients of linear expansion in~(\ref{eqN5.2}) and~(\ref{eqN5.3}) are actually spanned by two parameters $\sigma_e$ and $\sigma_s$ as
\begin{eqnarray}
\ \ \ \ \ \ \ \ \ \ \ \ \ \ \
\mu_s=\frac{A(t_0,T_s,T_e)}{A(t_0,T_x,T_e)}\sigma_s,&\quad &\mu_e=\frac{A(t_0,T_s,T_e)}{A(t_0,T_x,T_e)}\sigma_e,\nonumber\\
\nu_s=\frac{A(t_0,T_s,T_e)}{A(t_0,T_x,T_s)}\sigma_s,&\quad &\nu_e=\frac{A(t_0,T_s,T_e)}{A(t_0,T_x,T_s)}\sigma_e.
\label{sigmas}
\end{eqnarray}
We shall approximate linearly $G_{e,T_x}=w_1(y,x)^{-1}$ and $G_{s, T_x}=w_2(y,x)^{-1}$:
\begin{eqnarray}
G_{e,T_x}&\approx&\frac{A(t_0,T_s,T_e)}{A(t_0,T_x,T_e)}(1-\mu_e(y-R(t_0,T_x,T_e))-\mu_s(x-\tilde R(t_0,T_x,T_s))),
\label{eqN5.5}
\end{eqnarray}
\begin{eqnarray}
\tilde R(t_0,T_x,T_s)&=&{\mathbb E^{A(T_x,T_e)}}\left[R(t,T_x,T_s)\right],
\label{eqN5.6}
\end{eqnarray}
and
\begin{eqnarray}
G_{s,T_x}&\approx&\frac{A(t_0,T_s,T_e)}{A(t_0,T_x,T_s)}(1-\nu_e(y-\tilde R(t_0,T_x,T_e))-\nu_s(x-R(t_0,T_x,T_s))),
\label{eqN5.7}
\end{eqnarray}
\begin{eqnarray}
\tilde R(t_0,T_x,T_e)&=&{\mathbb E^{A(T_x,T_s)}}\left[R(t,T_x,T_e)\right],
\label{eqN5.8}
\end{eqnarray}
so that $G_{e,T_x}$ is $A(T_x,T_e)$-martingale and $G_{s,T_x}$ is $A(T_x,T_s)$-martingale.

Using Equations~(\ref{eqN5.5})-(\ref{eqN5.8}) we derive the following Lemma:
\begin{Lemma}
\label{gauss}
Under an assumption that 
the long and the short swap rates are approximately Gaussian the marginals of the joint distribution of the long and the short swap rates in $A(t,T_s,T_e)$-measure are  
\begin{eqnarray}
\ \ \ \ \ \ \ \ \
\phi_s(x)&\approx&{\rm PDF}^s_{R_s}(x)\left(1-\left(\nu_s+\nu_e\rho\frac{\Sigma_e}{\Sigma_s}\right)(x-R(t_0,T_x,T_s))\right),\nonumber \\
\phi_e(y)&\approx&{\rm PDF}^e_{R_e}(y)\left(1-\left(\mu_e+\mu_s\rho\frac{\Sigma_s}{\Sigma_e}\right)(y-R(t_0,T_x,T_e))\right).\nonumber\\
\label{mcf}
\end{eqnarray}
where $\Sigma_e,\Sigma_s$ are the volatilities of the long and the short swap rates:
\begin{equation}
\Sigma^2_e = {\mathbb E^{A(T_x,T_e)}}\left[\left(y-R(t_0,T_x,T_e)\right)^2\right], \quad{} 
\Sigma^2_s = {\mathbb E^{A(T_x,T_s)}}\left[\left(x-R(t_0,T_x,T_s)\right)^2\right].\nonumber
\end{equation}
\end{Lemma}
\noindent
{\bf Proof:} 
Under an assumption that 
the long and the short swap rates are approximately Gaussian we can project $y$ on to $x$ as:
\begin{eqnarray}
{\mathbb E^{A(T_x,T_s)}}[y|x] &= &{\mathbb E^{A(T_x,T_s)}}\left[R(T_x,T_x,T_e)|R(T_x,T_x,T_s)=x\right]\nonumber\\
&=&E^{A(T_x,T_s)}\left[R(T_x,T_x,T_e)\right] + \rho\frac{\Sigma_e}{\Sigma_s}(x-R(t_0,T_x,T_s)).
\end{eqnarray}
We can evaluate
\begin{eqnarray}
{\mathbb E^{A(T_x,T_s)}\left[G_{s,T_x}|R(T_x,T_x,T_s)=x\right]} &=&  \frac{A(t,T_s,T_e)}{A(t,T_x,T_s)}\left(1  -\left (\nu_s+\nu_e\rho\frac{\Sigma_e}{\Sigma_s}\right)(x-R(t,T_x,T_s))\right),\nonumber
\end{eqnarray}
which leads to the expression for the first marginal. Similarly we derive the expression for the second marginal.

\

Integrating $R(t,T_x,T_e)$ with respect to $\phi_e(y)$ and $R(t,T_x,T_s)$ with respect to $\phi_s(y)$ we derive
\begin{Lemma} 
\label{linfwd}
If the long and the short rates are approximately Gaussian then the linear approximation for the weights $w_1(y,x)$ and $w_2(y,x)$ leads to: 
\begin{eqnarray}
\hat R(t,T_x,T_e) &=& {\mathbb E^{A(T_s,T_e)}\left[R(T_x,T_x,T_e)\right]} = R(t_0,T_x,T_e) -  (\mu_e\Sigma_e + \mu_s\rho\Sigma_s)\Sigma_e,\nonumber\\
\hat R(t,T_x,T_s) &=& {\mathbb E^{A(T_s,T_e)}\left[R(T_x,T_x,T_s)\right]} = R(t_0,T_x,T_s) - (\nu_s\Sigma_s + \nu_e\rho\Sigma_e)\Sigma_s.
\label{hates}
\end{eqnarray}
\end{Lemma}

\
In practice, it may be convenient to calcuate $\tilde R(t_0,T_x,T_s)$ from (\ref{eqN5.6}) and $\tilde R(t_0,T_x,T_s)$ from (\ref{eqN5.8}).
This can be done by matching $w_1(y,x) G_{e,T_x} = 1$ and $w_2(y,x)G_{s,T_x}=1$ in the expectation relative to $A(T_s,T_e)$-measure.
\begin{Lemma} 
\label{linfwd}
If the long and the short rates are approximately Gaussian then the linear approximation for the weights $w_1(y,x)$ and $w_2(y,x)$ leads to: 
\begin{eqnarray}
\tilde R(t,T_x,T_e) &=& {\mathbb E^{A(T_x,T_e)}\left[R(T_x,T_x,T_e)\right]}\nonumber\\
&=& R(t_0,T_x,T_e) -  (\mu_e\Sigma_e + \mu_s\rho\Sigma_s)\Sigma_e+(\nu_e\Sigma_e + \nu_s\rho\Sigma_s)\Sigma_e,\nonumber\\
\tilde R(t,T_x,T_s) &=& {\mathbb E^{A(T_x,T_s)}\left[R(T_x,T_x,T_e)\right]}
\nonumber\\
&=& R(t_0,T_x,T_s) - (\nu_s\Sigma_s + \nu_e\rho\Sigma_e)\Sigma_s +(\mu_s\Sigma_s + \mu_e\rho\Sigma_e)\Sigma_s.
\label{tldes}
\end{eqnarray}
\end{Lemma}

\

Thus, in order to price a midcurve swaption we just need two extra parameters $\sigma_e$ and $\sigma_s$ from~(\ref{sigmas}). Together with the swap rates distributions  
${\rm PDF}^s_{R_s}(x)$, ${\rm PDF}^e_{R_e}(y)$ and the correlation between them, $\sigma_e$ and $\sigma_s$ uniquely determine the midcurve swaption price in the Gaussian copula model via~(\ref{eq5.22}),(\ref{eqN5.2}),(\ref{eqN5.3}), (\ref{mcf}) and (\ref{hates}).

\newpage 
\noindent
{\bf Log Linear Approximation:} The first order approximation does not immediately prevent weight coefficients $w_1(y,x)$ and $w_2(y,x)$ from going negative. This can be addressed by an exponential approximation for  $w_2(y,x)$: 
\begin{equation}
w_2(y,x)^{-1}=G_{s,T_x}=\alpha_s e^{-\nu_e \left(y -\tilde R(t,T_x,T_e)\right) -\nu_s \left(x - R(t,T_x,T_s)\right))},
\label{eqexp5.3}
\end{equation}
where
\begin{eqnarray}
\tilde R(t,T_x,T_e)&=&{\mathbb E^{A(T_x,T_s)}}\left[R(T_x,T_x,T_e)\right]
\label{lla1}
\end{eqnarray}
is used to underline that $G_{s,T_x}$ and $R(t,T_x,T_s)$ are $A(t,T_x,T_s)$-martingales.
The relation $w_1(y,x)-w_2(y,x)=1$ allows us to recover $w_1(y,x)$ from $w_2(y,x)$. We can evaluate the coefficient  $\alpha$ if we assume that 
the long and the short rates are approximately Gaussian. Let's project $y$ on to $x$ as:
\begin{eqnarray}
{\mathbb E^{A(T_x,T_s)}}[y|x] &= &{\mathbb E^{A(T_x,T_s)}}\left[R(T_x,T_x,T_e)|R(T_x,T_x,T_s)=x\right]\nonumber\\
&=&E^{A(T_x,T_s)}\left[R(T_x,T_x,T_e)\right] + \rho\frac{\Sigma_e}{\Sigma_s}(x-R(t_0,T_x,T_s)),
\end{eqnarray}
so that
\begin{eqnarray}
y -\tilde R(t,T_x,T_e) &=& \sqrt{1-\rho^2} \Sigma_e  z +   {\mathbb E^{A(T_x,T_s)}}[y - R(t,T_x,T_e)|x] \nonumber\\
&=&  \sqrt{1-\rho^2} \Sigma_e  z + \rho\frac{\Sigma_e}{\Sigma_s}(x-R(t,T_x,T_s)).
\end{eqnarray}
with $z\sim N(0,1)$. With this assumption we have
\begin{eqnarray}
E^{A(T_x,T_s)}\left[G_{s,T_x}\right] &=&\alpha_s e^{\nu^2_e(1-\rho^2)\Sigma^2_e/2}e^{\left(
(\nu_s \Sigma_s+\nu_e\rho\Sigma_e)^2/2\right)}\nonumber\\
&=&\alpha_s e^{\left(\nu^2_e\Sigma^2_e + 2\rho\nu_e\nu_s\Sigma_e\Sigma_s + \nu^2_s\Sigma^2_s\right)/2}\nonumber\\
&=&\frac{A(t,T_s,T_e)}{A(t,T_x,T_s)}.\nonumber\\
\alpha_s &=& \frac{A(t,T_s,T_e)}{A(t,T_x,T_s)}e^{-\left(\nu^2_e\Sigma^2_e + 2\rho\nu_e\nu_s\Sigma_e\Sigma_s + \nu^2_s\Sigma^2_s\right)/2}.
\label{alphas}
\end{eqnarray}
We can now evaluate
\begin{eqnarray}
{\mathbb E^{A(T_x,T_s)}\left[G_{s,T_x}|R(T_x,T_x,T_s)=x\right]} &=& \alpha_s {\mathbb E^{A(T_x,T_s)}}\left[e^{-\nu_e\sqrt{1-\rho^2}\Sigma_e z}\right] 
e^{-\left(\nu_s+\nu_e\rho\frac{\Sigma_e}{\Sigma_s}\right)(x-R(t_0,T_x,T_s))}\nonumber\\
&=& \frac{A(t,T_s,T_e)}{A(t,T_x,T_s)}e^{-\frac{\left(\rho\nu_e\Sigma_e+\nu_s\Sigma_s\right)^2}{2}}e^{-\left(\nu_s+\nu_e\rho\frac{\Sigma_e}{\Sigma_s}\right)(x-R(t,T_x,T_s))}\nonumber\\
\end{eqnarray}
This leads to the next result: 
\begin{Lemma}
If the long and the short rates are approximately Gaussian then the log linear approximation for the weights $w_2(y,x)$
\begin{eqnarray}
\phi_s(x)&\approx&{\rm PDF}^s_{R_s}(x)e^{-\frac{\left(\rho\nu_e\Sigma_e+\nu_s\Sigma_s\right)^2}{2}}e^{-\left(\nu_s+\nu_e\rho\frac{\Sigma_e}{\Sigma_s}\right)(x-R(t_0,T_x,T_s))},\nonumber \\
\hat R(t,T_x,T_s) &=&{\mathbb E^{A(T_s,T_e)}\left[R(T_x,T_x,T_s)\right]} = R(t_0,T_x,T_s) - (\nu_s\Sigma_s + \nu_e\rho\Sigma_e)\Sigma_s.
\end{eqnarray}
\label{phisexp} 
\end{Lemma}

Using the relation $w_1(y,x) - w_2(y,x) = 1$ we can reconstruct $G_{e,T_x}$ by numerical integration. To get less precise but more tractable formulae, instead, we evaluate $G_{e,T_x}$ as an exponential martingale:
\begin{equation}
w_1(y,x)^{-1}=G_{e,T_x}=\alpha_e e^{-\mu_e \left(y -R(t,T_x,T_e)\right) -\mu_s \left(x - \tilde R(t,T_x,T_s)\right)},\nonumber\\
\label{eqexp5.4}
\end{equation}
where
\begin{eqnarray}
\tilde R(t,T_x,T_s)&=&{\mathbb E^{A(T_x,T_e)}}\left[R(T_x,T_x,T_s)\right].
\label{lla2}
\end{eqnarray}
Similarly to (\ref{alphas}) we obtain
\begin{eqnarray}
\alpha_e &=& \frac{A(t,T_s,T_e)}{A(t,T_x,T_e)}e^{-\left(\mu^2_e\Sigma^2_e + 2\rho\mu_e\mu_s\Sigma_e\Sigma_s + \mu^2_s\Sigma^2_s\right)/2},
\end{eqnarray}
\begin{eqnarray}
{\mathbb E^{A(T_x,T_e)}\left[G_{e,T_x}|R(T_x,T_x,T_e)=y\right]} &=& \alpha_e {\mathbb E^{A(T_x,T_e)}}\left[e^{-\mu_s\sqrt{1-\rho^2}\Sigma_s z}\right] 
e^{-\left(\mu_e+\mu_s\rho\frac{\Sigma_s}{\Sigma_e}\right)(y-R(t_0,T_x,T_e))}\nonumber\\
&=& \frac{A(t,T_s,T_e)}{A(t,T_x,T_e)}e^{-\frac{\left(\rho\mu_s\Sigma_s+\mu_e\Sigma_e\right)^2}{2}}e^{-\left(\mu_e+\mu_s\rho\frac{\Sigma_s}{\Sigma_e}\right)(y-R(t,T_x,T_e))}.\nonumber\\
\end{eqnarray}
Similar to Lemma~\ref{phisexp} we derive:
\begin{Lemma} If the long and the short rates are approximately Gaussian then the exponential approximation for the weight $w_1(y,x)$ leads to: 
\begin{eqnarray}
\phi_e(y)&\approx&{\rm PDF}^e_{R_e}(y)e^{-\frac{\left(\rho\mu_s\Sigma_s+\mu_e\Sigma_e\right)^2}{2}}e^{-\left(\mu_e+\mu_s\rho\frac{\Sigma_s}{\Sigma_e}\right)(y-R(t_0,T_x,T_e))},\nonumber \\
\hat R(t,T_x,T_e) &=& {\mathbb E^{A(T_s,T_e)}\left[R(T_x,T_x,T_e)\right]} = R(t_0,T_x,T_e) - (\mu_e\Sigma_e + \mu_s\rho\Sigma_s)\Sigma_e.
\end{eqnarray}
\label{phieexp} 
\end{Lemma}
In the log linear approximation for the weights we can explicitly evaluate $\tilde R(t,T_x,T_s)$ from (\ref{lla2}) and  $\tilde R(t,T_x,T_e)$ from (\ref{lla1}) by observing that $w_1(y,x)$ and $w_2(y,x)$ are $A(t,T_s,T_e)$-martingales. 
We find, similarly to the linear case:
\begin{eqnarray}
\tilde R(t,T_x,T_s)&=& R(t_0,T_x,T_s) - (\nu_s\Sigma_s +\nu_e\rho\Sigma_e)\Sigma_s + (\mu_s\Sigma_s + \mu_e\rho\Sigma_e)\Sigma_s,\nonumber\\
w_1(y,x) &=& \frac{A(t,T_x,T_e)}{A(t,T_s,T_e)}e^{-\left(\mu^2_e\Sigma^2_e + 2\rho\mu_e\mu_s\Sigma_e\Sigma_s + \mu^2_s\Sigma^2_s\right)/2}\times\nonumber\\
&\times &
 e^{\mu_e \left(y -\hat R(t,T_x,T_e)\right) + \mu_s \left(x - \hat R(t,T_x,T_s)\right)},
\label{exprs}
\end{eqnarray}
and
\begin{eqnarray}
\tilde R(t,T_x,T_e)&=&R(t_0,T_x,T_e) -  (\mu_e\Sigma_e + \mu_s\rho\Sigma_s)\Sigma_e +(\nu_e\Sigma_e + \nu_s\rho\Sigma_s)\Sigma_e.\nonumber\\
w_2(y,x)&=&\frac{A(t,T_x,T_s)}{A(t,T_s,T_e)}e^{-\left(\nu^2_e\Sigma^2_e + 2\rho\nu_e\nu_s\Sigma_e\Sigma_s + \nu^2_s\Sigma^2_s\right)/2}\times\nonumber\\
&\times & e^{\nu_e \left(y -\hat R(t,T_x,T_e)\right) + \nu_s \left(x - \hat R(t,T_x,T_s)\right)}.
\label{expre}
\end{eqnarray}
With these exponential approximations for both $G_{s,T_x}$ and $G_{e,T_x}$ we shall chose parameters $\nu_e$, $\nu_s$, $\mu_e$, and $\mu_s$ to minimise
\begin{eqnarray}
{\mathbb E^{A(T_s,T_e)}\left[\left(w_1(y,x) - w_2(y,x)\right)^2\right]} && \nonumber\\
\end{eqnarray}
in $A(t,T_s,T_e)$-measure. 
Expanding up to the second order in volatilities $\Sigma_e$, $\Sigma_s$ we see that as soon as parameters $\mu_e$, $\mu_s$, $\nu_e$ and $\nu_s$ are related as in~(\ref{sigmas}) via:
\begin{eqnarray}
\ \ \ \ \ \ \ \ \ \ \ \ \ \ \
\mu_s=\frac{A(t_0,T_s,T_e)}{A(t_0,T_x,T_e)}\sigma_s,&\quad &\mu_e=\frac{A(t_0,T_s,T_e)}{A(t_0,T_x,T_e)}\sigma_e,\nonumber\\
\nu_s=\frac{A(t_0,T_s,T_e)}{A(t_0,T_x,T_s)}\sigma_s,&\quad &\nu_e=\frac{A(t_0,T_s,T_e)}{A(t_0,T_x,T_s)}\sigma_e,
\label{sigmasexp}
\end{eqnarray}
the variance of $w_1(y,x) - w_2(y,x)$ is zero up to the second order in $\Sigma_e$, $\Sigma_s$, i.e.:
\begin{equation}
w_1(y,x) - w_2(y,x) \approx 1 + \overline o(\Sigma^2_e,\Sigma^2_s).
\end{equation}
Again as in the linear case, in order to price a midcurve swaption we just need two extra parameters $\sigma_e$ and $\sigma_s$ from~(\ref{sigmasexp}). Together with the swap rates distributions  
${\rm PDF}^s_{R_s}(x)$ and ${\rm PDF}^e_{R_e}(y)$,  the correlation between them, $\sigma_e$ and $\sigma_s$ uniquely determine the midcurve swaption price in the Gaussian copula model via~(\ref{eq5.22}),(\ref{exprs}), (\ref{expre}),  from Lemma~\ref{phisexp} and  from Lemma~\ref{phieexp}.

\section{Estimating parameters $\sigma_e$ and $\sigma_s$ and some numerical results.}
Parameters $\sigma_e$ and $\sigma_s$ introduced in~(\ref{sigmas}) for the linear approximation and in~(\ref{sigmasexp}) for the log linear approximation, are related to the covariances between swap annuities' ratios and the swap rates via
\begin{eqnarray}
Cov^{A(T_x,T_e)}\langle\frac{A(T_x,T_s,T_e)}{A(T_x,T_x,T_e)} , R(T_x,T_x,T_e) \rangle&=& 
-\frac{A(T_x,T_s,T_e)^2}{A(T_x,T_x,T_e)^2}\left(\sigma_e\Sigma^2_e + \sigma_s\rho\Sigma_e\Sigma_s\right),\nonumber\\
Cov^{A(T_x,T_s)}\langle\frac{A(T_x,T_s,T_e)}{A(T_x,T_x,T_s)} , R(T_x,T_x,T_s) \rangle  &=& 
-\frac{A(t_0,T_s,T_e)^2}{A(t_0,T_x,T_s)^2} \left(\sigma_e\rho\Sigma_e\Sigma_s + \sigma_s\Sigma^2_s\right).
\label{covs}
\end{eqnarray}
The covariances on the left hand side of~(\ref{covs})  can be estimated either from the historical data or by using methods  suggested in~\cite{cedpit}. In~\cite{cedpit}, the authors used the expansions  for annuities in terms of swap rates for modeling CMS claims dependence on the volatilities and the correlations of the swap rates beyond the smile of the underlying CMS rate. The methods~\cite{cedpit} can be immediately adapted to our case. 
Using the non-linear annuity mapping from~\cite{cedpit} with a single stochastic driver, first we link all of the swap rates $R(t, T_x,T_i)$, $i=1, \dots e$, which fix at $T_x$ to the stochastic driver $Y\sim N(0,1)$ behind the long swap rate $R(T_x,T_x,T_e)$:
\begin{equation}
1+\tau_i R_i(T_x,T_x,T_i) \approx (1+\tau_i R_i) e^{\mu_i +\nu_i Y}, \quad R_i = {\mathbb E^{A(T_x,T_i)}}\left[ R(t,T_x,T_i)\right],
\end{equation}
with $\nu_i$ given by the variance  $\Sigma^2_i$ of the swap rate $R(T_x,T_x,T_i)$ and its correlation $\rho_{e,i}$ with the long swap rate $R(T_x,T_x,T_e)$ via
\begin{equation}
\nu_i \approx \frac{\tau_i}{1+\tau_i R_i}\rho_{e,i}\Sigma_i,
\end{equation} 
and $\mu_j$ determined (usig $T_x$-forward measure) recursively from
\begin{eqnarray}
A_j &=& {\mathbb E^{T_x}}\left[ A(T_x,T_x,T_j)\right]\nonumber\\
&=&\sum^j_{i=1} \tau_i exp\left(-\sum^j_{k=i}\mu_k +\frac{1}{2} \left(\sum^j_{k=i} \nu_k\right)^2\right)\prod^j_{k=i}\frac{1}{1+\tau_k R_k}.
\end{eqnarray}
Within this parameterisation the annuities $A(T_x,T_x,T_e)$ and $A(T_x,T_x,T_s)$ are given by~\cite{cedpit}:
\begin{eqnarray}
A(T_x,T_x,T_e) &=& \sum^e_{i=1} \tau_i exp\left(-\sum^e_{k=i}\left(\mu_k +\nu_kY\right)\right)\prod^e_{k=i}\frac{1}{1+\tau_k R_k},\nonumber\\
A(T_x,T_x,T_s) &=& \sum^s_{i=1} \tau_i exp\left(-\sum^s_{k=i}\left(\mu_k +\nu_kY\right)\right)\prod^s_{k=i}\frac{1}{1+\tau_k R_k}.
\end{eqnarray}
Linearising the ratio $(A(T_x,T_x,T_e) - A(T_x,T_x,T_s))/A(T_x,T_x,T_e)$ with respect to $Y$, we estimate the left hand side of the first equation in~(\ref{covs}). Repeating the same procedure for the short rate $R(T_x,T_x,T_s)$ and the corresponding driver $X\sim N(0,1)$, we estimate the left hand side of the second equation in~(\ref{covs}).
 
We conducted a numerical experiment by Monte Carlo integration of the two dimensional Gaussian copula to estimate the implied correlation of the midcurve swaption.  To simplify the calculation of the coefficients $\sigma_e$ and $\sigma_s$ along the lines of~\cite{cedpit} we study the case of $1y\to1y1y$ midcurve, where the swaption expires in one year, and the holder has the right to enter into a one year swap starting one year after the expiry (equivalently, starting two years from now). We assume that all of the swap rates' payment frequencies are annual, i.e. all of the accrual fractions $\tau_i =1$. 
We set the short 1y swap rate at $2.631\%$ with the normal volatility of 60.00 bps and the long 2y rate at $2.2347\%$ with the normal volatility of 64.18 bps. We choose the correlation between the long and the short swap rates to be $80\%$. Calculation~\cite{cedpit}, as described in the beginning of the section for the $1y\to 1y1y$ midcurve with annually paid coupons, is easier than in the generic case, and we can estimate indicative levels for parameters $\sigma_e$ and $\sigma_s$ using analytical expressions for the annuities  in terms of just the long and the short swap rates. We found
$\sigma_e$  to be close to 2.0 and $\sigma_s$ to be close to -1.0.  

\begin{figure}[h!]
\centering
\begin{minipage}[b]{1.0\textwidth}
 \includegraphics[width=\textwidth, height=75mm]{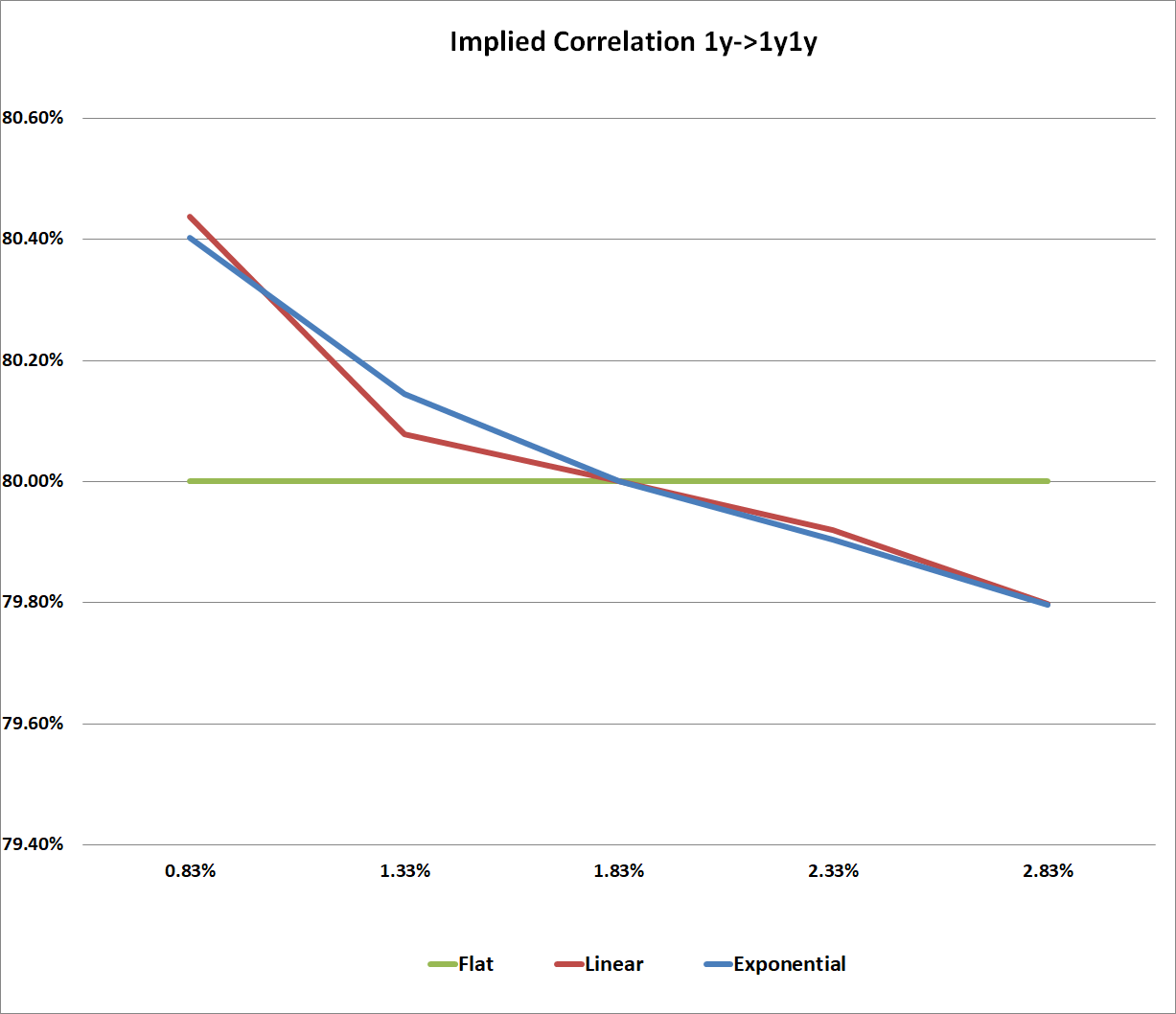}
\caption{\footnotesize{Midcurve 1y$\to$1y1y, forward swap (ATM) rate $1.8294\% $, the short swap rate normal vol 60 bps, the long swap rate normal vol 64.18 bps, flat vol smile, the swap rates correlation $80\%$ and $\sigma_e=2$, $\sigma_s = -1$.}}
\label{ImplCorr}
\end{minipage}
\end{figure}

Under the assumption of no volatility smile, we implied the correlation by strike from the prices of midcurve swaptions evaluated by Monte Carlo integration. 
From Figure~\ref{ImplCorr}, we see that the stochastic annuity ratio assumption introduces a skew into the midcurve implied correlation even in the case of flat implied volatilities of the long and the short swap rates.

\section*{Conclusion}

We developed a consistent model for midcurve swaption pricing which explicitly accounts for stochasticity of the ratios of the annuities. It gives a handle on the correlation skew which is typically risk managed via the correlation-by-strike. The latter approach is not arbitrage free.

Our paper shares a common idea with~\cite{cedpit}, and, thus, depending on the size of the book the model presented here can be used for trading a small number of midcurve products and understanding their correlation risk in terms of linear regression coefficients $\sigma_e$ and $\sigma_s$, or the model can be used for risk managing large books of swaptions and CMS products via full projection of  the correlation risk on all of the swap rates' volatilities and all of their pairwise correlations.

\

\noindent
{\bf Email address:} kostyafeldman@gmail.com.

\end{document}